\documentclass[aps,prb,twocolumn,superscriptaddress,showpacs]{revtex4-1}

\usepackage[dvips]{graphicx}
\usepackage{amssymb,amsmath,amsfonts}

\begin{document}


\title{Energy gap tuning in graphene on hexagonal boron nitride bilayer system}



\author{J. S\l awi\'{n}ska}
\email{jagoda.slawinska@gmail.com} 
\affiliation{Theoretical Physics Department II}
\author{I. Zasada}
\author{Z. Klusek}
\affiliation{Solid State Physics Department,\\
 University of Lodz, Pomorska 149/153, 90-236 Lodz, Poland}

\vspace{.15in}

\begin{abstract}
We use a tight binding approach and density functional theory calculations to study the band structure of graphene/hexagonal boron nitride bilayer system in the most stable configuration. We show that an electric field applied in the direction perpendicular to the layers significantly modifies the electronic structure of the whole system, including shifts, anticrossing and other deformations of bands, which can allow to control the value of the energy gap. It is shown that band structure of biased system may be tailored for specific requirements of nanoelectronics applications. The carriers' mobilities are expected to be higher than in the bilayer graphene devices.
\end{abstract}


\maketitle

\section{Introduction}
Graphene, a one atom thick graphitic carbon, since its fabrication in 2004\cite{novoselov_science} has attracted a huge attention due to its exceptional electronic properties\cite{review} and currently is a subject of extensive research in condensed matter physics. Due to extremely high carrier mobility graphene is considered to be an ideal material for carbon-based nanoelectronics. The production of large enough graphene sheets makes it a promising candidate for direct applications, especially as a channel material in field effect transistors (FETs).

Achieving large on/off current ratios is essential for logic devices. It has been demonstrated that semimetallic single layer graphene (SLG) is not a suitable choice for FETs constructions,\cite{transistors} since only an on/off current ratio of about 4 has been observed.\cite{nanoletters} Although graphene nanoribbons field effect transistors (GNRFETs) have been produced, exhibiting excellent properties\cite{gnr1, gnr2}, mass production of GNRs-based devices is beyond abilities of current litography techniques.\cite{nilsson}

The opening and external tuning of energy gap between valence and conduction bands in Bernal stacking bilayer graphene (BLG)\cite{mccann, biased} reveals that it can have the largest potential for logic applications. It has been demonstrated experimentally using angle-resolved photoemission spectroscopy (ARPES) that energy gap tuning can be achieved in epitaxial graphene.\cite{ohta} Charge transfer from SiC substrate and additional doping by potassium atoms on top of the graphene sample induce effective external field controlled by changes of dopants amount. However, recent experimental studies on very large number of samples produced in an integrated technology\cite{kedzierski} show that switching off the conduction to desirable level is still a very difficult task in epitaxial graphene.

Energy gap tuning by electric field effect has been also detected in exfoliated bilayer graphene deposited on SiO$_{\textrm{2}}$/Si wafer controlled by using double gate device.\cite{oostinga, castro_prl} The optical measurments \cite{zhang, mak} have confirmed predictions from tight binding calculations that energy gap saturates and does not achieve more than $\gamma_{1}\sim0.35$~eV.\cite{hongkimin} A recent study of carrier transport in state-of-the-art bilayer graphene FET device\cite{nanoletters} indicates that a transport band gap of $>$130 meV and on/off current ratio of around 100 at room temperature is observed. 

In bilayer graphene, in contrast to single layer sheet, the Dirac fermions are massive, with an effective mass $m^{*}\approx\,0.03\,m_{\textrm{e}}$ (where $m_{\textrm{e}}$ is the bare electron mass).\cite{conference} Since the exceptional conical dispersion is not then preserved, an important question arises whether it is possible to open a band gap and still keep the extraordinary electronic properties of SLG. Several proposals of band gap engineering have been recently given, including surface bonding (graphane), isoelectronic co-doping with boron and nitrogen atoms and alternating electrical or chemical environment (by using Li and F atoms).\cite{appl} It has been demonstrated, however, that these systems may not offer higher carriers' mobilities than BLG-based semiconductors. We believe that the crucial issue of band gap tuning and at the same time preserving the exceptional linear dispersion can be achieved by using suitable substrates. 

We reintroduce here the idea of graphene on hexagonal boron nitride (h-BN) given in Ref.~\onlinecite{hbn}. In contrast to graphene, the isoelectronic h-BN crystal with strong, partially ionic sp$^{\textrm{2}}$ hybrydized in-plane bonding between boron and nitrogen atoms,\cite{ksiazka} has a wide energy gap resulting from a significant difference of electronegativities between onsite atoms. It emerges in a natural way that structural and band structures similarities between these two materials can lead to amazing physical properties of a combined system. 

It has been shown\cite{hbn} that the most stable configuration of a graphene on h-BN substrate has one carbon atom on top of a boron atom while the other one sits above a center of boron nitride ring. The density functional theory (DFT) calculations indicated the opening of a gap of 53 meV at the Dirac points resulting from the graphene-substrate interaction. On the other hand, as indicated in Ref.~\onlinecite{semenoff}, one can calculate the band structures of graphene and hexagonal boron nitride diagonalizing a hamiltonian of nearly the same form. This similarity allows to calculate conduction and valence bands of graphene/h-BN bilayer system in almost the same manner as for graphite,\cite{wallace} BLG and FLG.\cite{partoens4,partoens2,partoens3} The extension to the case of boron nitride sheet in place of carbon bottom layer is straightforward and perpendicular external fields can be easily taken into account. 

In the present paper the electronic properties of a biased graphene/h-BN system are studied within a tight binding (TB) approach verified by \textit{ab initio} calculations. The band structure is calculated both in the presence and in the absence of a perpendicular electric field. We study how the gap can be externally tuned by the electric field effect. The advantage of the present approach is that substrate, directly incorporated into the model, is an integral part of system which is indispensable in modelling the band structure.
It is shown that graphene on h-BN layer can be an alternative system with externally tunable energy gap. 

The paper is organized as follows. In Sec.~II we discuss the band structure evolution from bilayer graphene towards the graphene on hexagonal boron nitride in the framework of the nearest-neighbour tight binding formalism. Since the hopping parameters needed in the tight binding description are taken from experimental or \textit{ab initio} calculations data (TB-LDA,\cite{lda} TB-GW\cite{tb}), the fit of TB energy bands to DFT results is needed and presented in Sec.~III. The final remarks and perspectives are reported in Sec.~IV.

\section{Tight binding calculations}

\begin{figure*}
\includegraphics{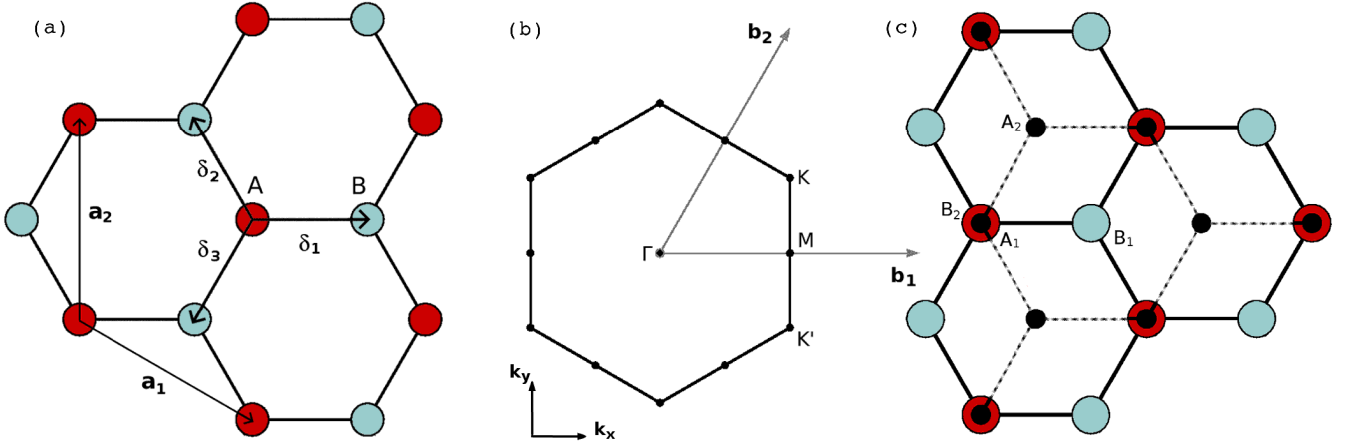}
\caption{\label{sieci}(Color online) (a) Lattice structure of graphene (or h-BN) constructed by two interpenetrating triangular sublattices, (b) corresponding Brillouin zone and (c) lattice structure of graphene (black points) on h-BN. Boron atoms are denoted as red (darker) circles, nitrogen atoms as blue (lighter) ones.}
\end{figure*}

In periodic table of elements carbon is placed between boron and nitrogen, thus they differ only in the number of $2p$ electrons. Since the honeycomb structure of graphite originates from $sp^{2}$ hybridized in-plane bonding, the proximity of electron energies of $2s$ and $2p$ states in carbon, boron and nitrogen\cite{ksiazka} results in formation of similar structure in the case of boron nitride. Graphene and h-BN have identical lattices, but, in contrast to zero energy gap in graphene, the gap in h-BN is of the order of 5$\;$eV depending on the number of layers. The difference between two materials can be explained within TB approach.\cite{semenoff}

Both crystals have two dimensional hexagonal structure shown in Fig.~\ref{sieci}\,(a). A unit cell contains two sites, one of type A and the second one of type B. The linear combinations of lattice vectors:
\begin{equation}\vec{a}_{1}=a(\frac{\sqrt{3}}{2},-\frac{1}{2}),\quad\vec{a}_{2}=a(0,1)\end{equation}
generate only A sites while sites in B sublattice are generated by $n_{1}\vec{a}_{1}+n_{2}\vec{a}_{2}+\vec{\delta}$, where $\vec{\delta}$ has to be choosen as one of the three nearest neighbours vectors: 
\begin{displaymath}\vec{\delta}_{1}=a(\frac{1}{\sqrt{3}},0), \quad\vec{\delta}_{2}=a(-\frac{1}{2\sqrt{3}},\frac{1}{2})\end{displaymath}
\begin{equation}\vec{\delta}_{3}=a(-\frac{1}{2\sqrt{3}},-\frac{1}{2})\end{equation}
The quantity $a$ stands for a lattice constant. The corresponding Brillouin zone (see Fig.~\ref{sieci}\,(b)) is a hexagon in reciprocal space with basis vectors given by:
\begin{equation}
\vec{b}_{1}=\frac{4\pi}{\sqrt{3}a}(1,0),
\quad\vec{b}_{2} =\frac{4\pi}{\sqrt{3}a}(\frac{1}{2},\frac{\sqrt{3}}{2})\end{equation}
Two inequivalent corners of the Brillouin zone $K$ and $K'$ can be chosen as follows:
\begin{equation}\vec{K}=\frac{4\pi}{\sqrt{3}a}(\frac{1}{2},\frac{1}{2\sqrt{3}}),\quad\vec{K'}=\frac{4\pi}{\sqrt{3}a}(\frac{1}{2},-\frac{1}{2\sqrt{3}})\end{equation}
Although there is a slight difference between lattice constants of graphene and hexagonal boron nitride, we take the same value for~both materials.\cite{hbn}

We present a simple model of graphene on hexagonal boron nitride in Fig.~\ref{sieci}\,(c). The top layer consists of carbon atoms, while the bottom layer is a hexagonal boron nitride sheet. As indicated in Ref.~\onlinecite{hbn}, it is the lowest energy orientation of the graphene layer with respect to the h-BN with the equilibrium distance between the sheets $c=3.22$\;\AA. A two dimensional Brillouin zone is shown in Fig.~1\,(b). One can regain the AB stacked bilayer graphene replacing boron and nitrogen by carbon atoms and substituting interlayer distance by $c=3.35$\AA.

Due to $sp^{2}$ hybrydization the resulted three $\sigma$ bonds are located in the graphene plane while two overlapping $2p_{z}$ orbitals form $\pi$ bond perpendicular to the sheet. Therefore $\sigma$ and $\pi$ orbitals can be analyzed independently. We restrict our considerations only to $\pi$ energy bands because they are responsible for unique properties of graphene. Two Bloch functions, constructed from atomic orbitals for two inequivalent carbon atoms A and B, provide the basis functions for single layer graphene. For bilayer system we assume a basis as $2p_{z}$ orbitals of respective atoms with Bloch factors:
\begin{subequations}\label{baza2}\begin{eqnarray}
\Phi^{\alpha_{1}}_{\vec{k}}(\vec{r})=\frac{1}{\sqrt{N}}\sum_{\vec{R}_{\alpha_{1}}}e^{i\vec{k}\vec{R}_{\alpha_{1}}}\phi_{\alpha_{1}}(\vec{r}-\vec{R}_{\alpha_{1}})
\\\Phi^{\beta_{1}}_{\vec{k}}(\vec{r})=\frac{1}{\sqrt{N}}\sum_{\vec{R}_{\beta_{1}}}e^{i\vec{k}\vec{R}_{\beta_{1}}}\phi_{\beta_{1}}(\vec{r}-\vec{R}_{\beta_{1}})
\\\Phi^{\alpha_{2}}_{\vec{k}}(\vec{r})=\frac{1}{\sqrt{N}}\sum_{\vec{R}_{\alpha_{2}}}e^{i\vec{k}\vec{R}_{\alpha_{2}}}\phi_{\alpha_{2}}(\vec{r}-\vec{R}_{\alpha_{2}})
\\\Phi^{\beta_{2}}_{\vec{k}}(\vec{r})=\frac{1}{\sqrt{N}}\sum_{\vec{R}_{\beta_{2}}}e^{i\vec{k}\vec{R}_{\beta_{2}}}\phi_{\beta_{2}}(\vec{r}-\vec{R}_{\beta_{2}})
\end{eqnarray}\end{subequations}
where $\vec{R}_{\alpha_{1}}$, $\vec{R}_{\beta_{1}}$ and $\vec{R}_{\alpha_{2}}$, $\vec{R}_{\beta_{2}}$ are the positions of inequivalent atoms in the first and second layer, while $\phi_{\alpha_{1}}$, $\phi_{\beta_{1}}$ and $\phi_{\alpha_{2}}$, $\phi_{\beta_{2}}$ are $2p_{z}$ orbitals in respective atoms. We use atomic valence orbitals and orbitals with close values of energies\cite{kolos} (\ref{baza2}) to construct the eigenfunctions:
\begin{equation}\Psi_{\vec{k}}(\vec{r})=\sum_{i=1}^{2}\,\biggl(c^{\alpha_{i}}_{\vec{k}}\Phi^{\alpha_{i}}_{\vec{k}}(\vec{r})+c^{\beta_{i}}_{\vec{k}}\Phi^{\beta_{i}}_{\vec{k}}(\vec{r})\biggr)\end{equation}

According to the procedure given in Ref.~\onlinecite{nanotubes} we introduce the following quantities representing the energies of electrons of $2p$ level in the crystal: 
\begin{subequations}\label{energie}\begin{eqnarray}\epsilon^{\alpha_{i}}_{2p_{z}}=\int \mathrm{d}\vec{r}\;\phi^{*}_{\alpha_{i}}(\vec{r}-\vec{R}_{\alpha_{i}})H\phi_{\alpha_{i}}(\vec{r}-\vec{R}_{\alpha_{i}})\\
\epsilon^{\beta_{i}}_{2p_{z}}=\int \mathrm{d}\vec{r}\; \phi^{*}_{\beta_{i}}(\vec{r}-\vec{R}_{\beta_{i}})H\phi_{\beta_{i}}(\vec{r}-\vec{R}_{\beta_{i}})\end{eqnarray}\end{subequations}
and in-plane and an out-of-plane hopping parameters:
\begin{equation}\gamma_{0}'=\int \mathrm{d}\vec{r}\;\phi^{*}_{\alpha_{1}}(\vec{r}-\vec{R}_{\alpha_{1}})H\phi_{\beta_{1}}(\vec{r}-\vec{R}_{\alpha_{1}}-\vec{\delta})\end{equation}
\begin{equation}\gamma_{0}=\int \mathrm{d}\vec{r}\;\phi^{*}_{\beta_{2}}(\vec{r}-\vec{R}_{\beta_{2}})H\phi_{\alpha_{2}}(\vec{r}-\vec{R}_{\beta_{2}}+\vec{\delta})\end{equation}
\begin{equation}
\gamma_{1}=\int \mathrm{d}\vec{r}\;\phi^{*}_{\alpha_{1}}(\vec{r}-\vec{R}_{\alpha_{1}})H\phi_{\beta_{2}}(\vec{r}-\vec{R}_{\alpha_{1}}-\vec{c})
\end{equation}
where $H$ is a hamiltonian of the system, $\vec{c}=c\,\vec{e}_{z}$ and only the  nearest neighbours interactions are assumed in the system. 
We construct then a hamiltonian in the basis (\ref{baza2}):
\begin{equation}H=\left( \begin{array}{cccc}\vspace{0.1in}
\epsilon^{\alpha_{1}}_{2p_{z}}& \gamma_{0}'f(\vec{k})&0&\gamma_{1}\\\vspace{0.1in}
\gamma_{0}'f(\vec{k})^{*} & \epsilon^{\beta_{1}}_{2p_{z}}&0&0\\\vspace{0.1in}
0& 0&\epsilon^{\alpha_{2}}_{2p_{z}}&\gamma_{0}f(\vec{k})\\
\gamma_{1}& 0&\gamma_{0}f(\vec{k})^{*}&\epsilon^{\beta_{2}}_{2p_{z}} \end{array} \right)\label{ham}\end{equation}
with 
\begin{equation}f(k_{x},k_{y})=\sum_{i=1}^{3}e^{i\vec{k}\vec{\delta}_{i}}=e^{iak_{x}/\sqrt{3}}+2e^{-iak_{x}/2\sqrt{3}}\cos(\frac{1}{2}ak_{y})\label{funkcja}\end{equation}
and we define the overlapping matrix in the form:
\begin{equation}S=\left( \begin{array}{cccc}\vspace{0.1in}
1& s'_{0}f(\vec{k})&0&s_{1}\\\vspace{0.1in}
s'_{0}f(\vec{k})^{*} &1&0&0\\\vspace{0.1in}
0& 0&1&s_{0}f(\vec{k})\\
s_{1}& 0&s_{0}f(\vec{k})^{*}&1 \end{array} \right)\label{overlap}\end{equation}
where $s_{0}$, $s'_{0}$ and $s_{1}$ denote overlapping integrals which are assumed to be small. In fact, since we study the band structures in the vicinity of the Brillouin zone corners, we can neglect all of them.\cite{partoens4}

The eigenvalue problem is simple in case of monolayer system and we start with providing a short outline of the obtained results.

\subsection{Graphene and boron nitride monolayers - an outline}
 
In case of graphene monolayer the hamiltonian (\ref{ham}) reduces to right bottom block with the orbital energies of the $2p_{z}$ level in carbon lattice $\epsilon^{\alpha}_{2p_{z}}=\epsilon^{\beta}_{2p_{z}}=\epsilon^{c}_{2p_{z}}$. We have the solutions:
\begin{equation}E_{\pm}(\vec{k})=\epsilon^{c}_{2p_{z}}\pm\gamma_{0}|f(\vec{k})|\end{equation}
where $\pm$ signs define two branches of the energy dispersion curves and $E_{-}(k)$, $E_{+}(k)$ are called bonding $\pi$ and antibonding $\pi^{*}$ energy bands, respectively.
It is seen from definition (\ref{funkcja}) that at the $K$ and $K'$ points it holds:
\begin{equation}|f|=0\label{warunek}\end{equation}
and a zero band gap is easily recognized. It means that the electrons in two sublattices do not interfere if they are at the Brillouin zone corners and it is a direct consequence of the symmetry of the system. It is worthwhile to remind that this does not change if hopping between non nearest neighbours is included in the model.

The hamiltonian (\ref{ham}) reduces to left top block for the case of boron nitride monolayer. The solutions can be written in a simple form:
\begin{equation}E_{\pm}(\vec{k})=\frac{1}{2}\biggr(\epsilon^{\alpha}_{2p_{z}}+\epsilon^{\beta}_{2p_{z}}\pm\sqrt{E_{g}^{2}+4\gamma_{0}'\,^{2}|f(\vec{k})|^{2}}\biggl)\end{equation}
with
\begin{equation}E_{g}=|\epsilon^{\alpha}_{2p_{z}}-\epsilon^{\beta}_{2p_{z}}|\end{equation}
defining the band gap at $K$ point. 
Then, in the corners of the Brillouin zone we have:

\begin{equation}E_{\pm}=\frac{1}{2}(\epsilon^{\alpha}_{2p_{z}}+\epsilon^{\beta}_{2p_{z}}\pm E_{g})\end{equation}

In this case the symmetry between A and B sublattices is broken because the energies of the onsite atoms are not the same representing boron
and nitrogen, respectively. The band structure is similar, but at the $K$ ($K'$) point a gap with the magnitude $E_{g}$ opens. In the first approximation, it emerges that energies $\epsilon^{\alpha}_{2p_{z}}$ and $\epsilon^{\beta}_{2p_{z}}$ could be taken close to energies of N$2p_{z}$ and B$2p_{z}$ states in respective atoms. However, for such choice, the value of $E_{g}$ would be about 0.2 eV while in the boron nitride a wider energy gap (5.9 eV) is observed. 

This considerable separation between valence and conduction bands is explained by difference in electronegativities of boron and nitrogen which is responsible for a negative charge transfer in the direction of nitrogen. Then the covalent bond gains partially ionic character which produces a Madelung energy. By means of DFT calculations it has been found that for a freestanding h-BN sheet about 0.56
electrons are transferred from B to N.\cite{blaha} This yields a sizable Madelung contribution to the lattice stability of 2.4 eV per 1x1 unit cell. 
The distinguishability of boron and nitrogen leads to the insulator behaviour. The valence band is mainly constituted by the nitrogen sublattice while the conduction band by the boron sublattice. This property is reflected in a density of states obtained by the \textit{ab initio} calculations.\cite{turcy}

\subsection{Bilayer graphene versus graphene on hexagonal boron nitride}
\begin{figure*}
\includegraphics{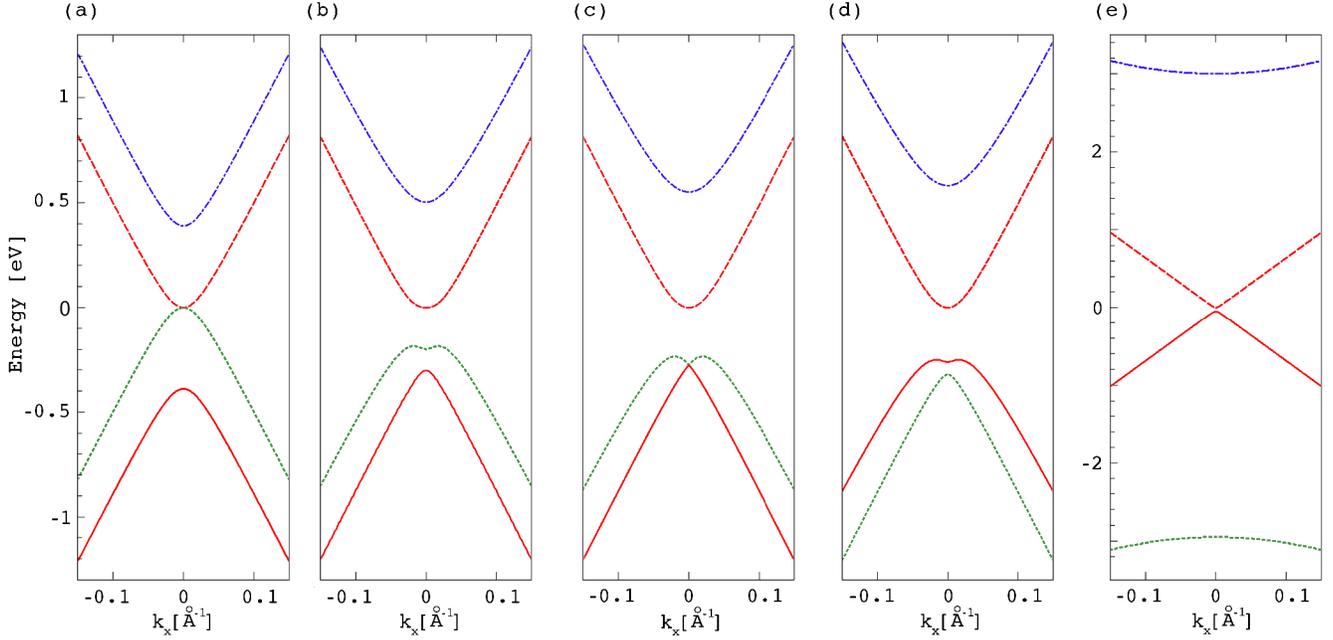}
\caption{\label{wykresy}(Color online) Band structure near $K$ (or $K'$) point for (a) graphene bilayer and (e) graphene on h-BN with $\epsilon^{\alpha_{1}}$=-$\epsilon^{\beta_{1}}=2.95$ eV. Graphs (b), (c) and (d) show the band structures for bilayer systems with different values of energy parameters $\epsilon^{\alpha_{1}}_{2p_{z}}$ and $\epsilon^{\beta_{1}}_{2p_{z}}$ describing the evolution of band structure from graphene bilayer towards graphene on h-BN. The onsite energies sets used for these three graphs: (b) $\epsilon^{\alpha_{1}}\,=\,-\epsilon^{\beta_{1}}=0.2$ eV, (c) $\epsilon^{\alpha_{1}}=-\epsilon^{\beta_{1}}=|\gamma_{1}|/\sqrt{2}$ and (d) $\epsilon^{\alpha_{1}}$=-$\epsilon^{\beta_{1}}=0.32$ eV.
The hopping parameters of graphite $\gamma_{1}$ and $\gamma_{0}$ known from experimental data\cite{experiment} fitted into the SWMcC model\cite{slonczewski, mcc} and $\gamma_{0}'=\gamma_{0}$ have been chosen for presented calculations. Blue dotted-dashed line denotes a band with the value $E_{1}$ (eq.\ref{e1}) at the Dirac points; red dashed line is assigned to the value $E_{2}$ (eq.\ref{e2}); olive dotted to the value $E_{4}$ (eq.\ref{e4}) and red solid to the value $E_{3}$ (eq.\ref{e3}).}
\end{figure*} 

A secular problem for bilayer systems leads to a quartic equation, which can be solved by radicals. For given parameters one can provide the solutions explicitly as a function of $\vec{k}$. At $K$ and $K'$ points the condition (\ref{warunek}) holds, thus we can express the eigenvalues of (\ref{ham}) for arbitrary parameters:
\begin{equation}E_{1}=\frac{1}{2}(\epsilon^{\alpha_{1}}_{2p_{z}}+\epsilon^{\alpha_{2}}_{2p_{z}})+\sqrt{\biggl(\frac{\epsilon^{\alpha_{1}}_{2p_{z}}+\epsilon^{\alpha_{2}}_{2p_{z}}}{2}\biggr)^{2}+\gamma_{1}^{2}-\epsilon^{\alpha_{1}}_{2p_{z}}\epsilon^{\alpha_{2}}_{2p_{z}}}\label{e1}\end{equation}\begin{equation}E_{2}=\epsilon^{\alpha_{2}}_{2p_{z}}\label{e2}\end{equation}\label{e3}\begin{equation}E_{3}=\frac{1}{2}(\epsilon^{\alpha_{1}}_{2p_{z}}+\epsilon^{\alpha_{2}}_{2p_{z}})-\sqrt{\biggl(\frac{\epsilon^{\alpha_{1}}_{2p_{z}}+\epsilon^{\alpha_{2}}_{2p_{z}}}{2}\biggr)^{2}+\gamma_{1}^{2}-\epsilon^{\alpha_{1}}_{2p_{z}}\epsilon^{\alpha_{2}}_{2p_{z}}}\label{e3}\end{equation}\begin{equation}E_{4}=\epsilon^{\beta_{1}}_{2p_{z}}\label{e4}\end{equation}

In Fig. \ref{wykresy} we plot the band structures of the system described by the general hamiltonian (\ref{ham}) for various choices of (\ref{energie}) and fixed hopping parameters. For simplicity we assume $\epsilon^{\alpha_{2}}=0$ and $\epsilon^{\alpha_{1}}=-\epsilon^{\beta_{1}}$. We start with the freestanding BLG (Fig. \ref{wykresy}\,(a), $\epsilon^{\alpha_{1}}=\epsilon^{\beta_{1}}=0$). A difference between onsite energies (Fig. \ref{wykresy}\,(b)) generates the 'Mexican hat' structure in the valence band which is also observed in bilayer graphene with substrate induced asymmetry.\cite{mucha} Further increasing of the asymmetry value leads to the touching of two valence bands at the $K$ and $K'$ points (Fig. \ref{wykresy}\,(c)). For the values of $\epsilon^{\alpha_{1}}$ larger than $|\gamma_{1}|/\sqrt{2}$ \,(Fig. \ref{wykresy}\,(d)) the bands assigned to the eigenvalues $E_{3}$ and $E_{4}$ change their order, thus in the limit of graphene on h-BN a lower valence band gains N$\,2p_{z}$ character (Fig. \ref{wykresy}\,(e)).

The band structure of graphene/h-BN system presented in Fig. \ref{wykresy}\,(e) has been obtained for the values of $\epsilon^{\alpha_{1}}=-\epsilon^{\beta_{1}}$ chosen to assure a proper energy gap for the case of h-BN monolayer described in the previous section. In consequence, a gap between valence and conduction bands is about 50~meV  which is of the same order as the one obtained from DFT calculations for graphene on four layers of h-BN substrate.\cite{hbn} In the vicinity of the Brillouin zone corners the band structure possesses nearly carbon character indicating a weak interaction between the layers.

In the present considerations, we have shown that evolution from BLG towards the graphene on h-BN system is related to changes of the onsite asymmetry in the bottom layer. 

\subsection{Bilayer system in an external electric field}
\begin{figure*}
\includegraphics{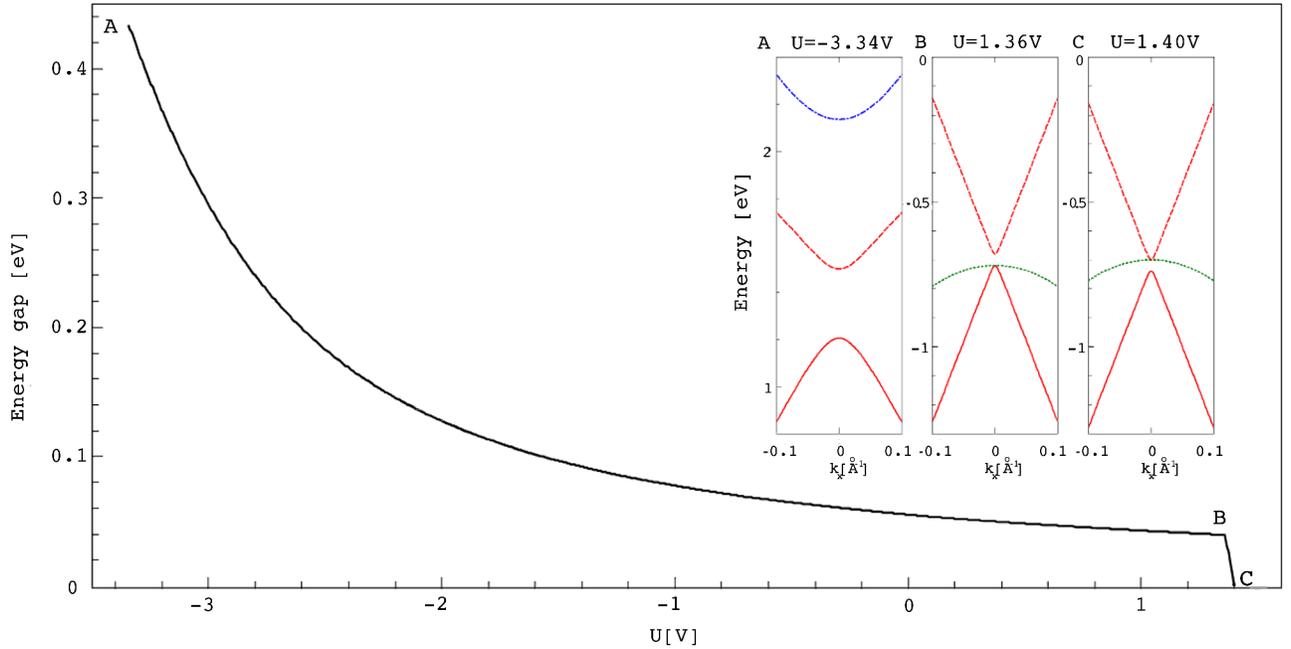}
\caption{\label{charakterystyka}(Color online) Variation of the energy gap as a function of energy difference between two layers. The insets show the band structures in three characteristic points indicated on the gap variation curve. The TB parameters are taken from Table \ref{parametry}.}
\end{figure*}
\begin{figure*}
\includegraphics{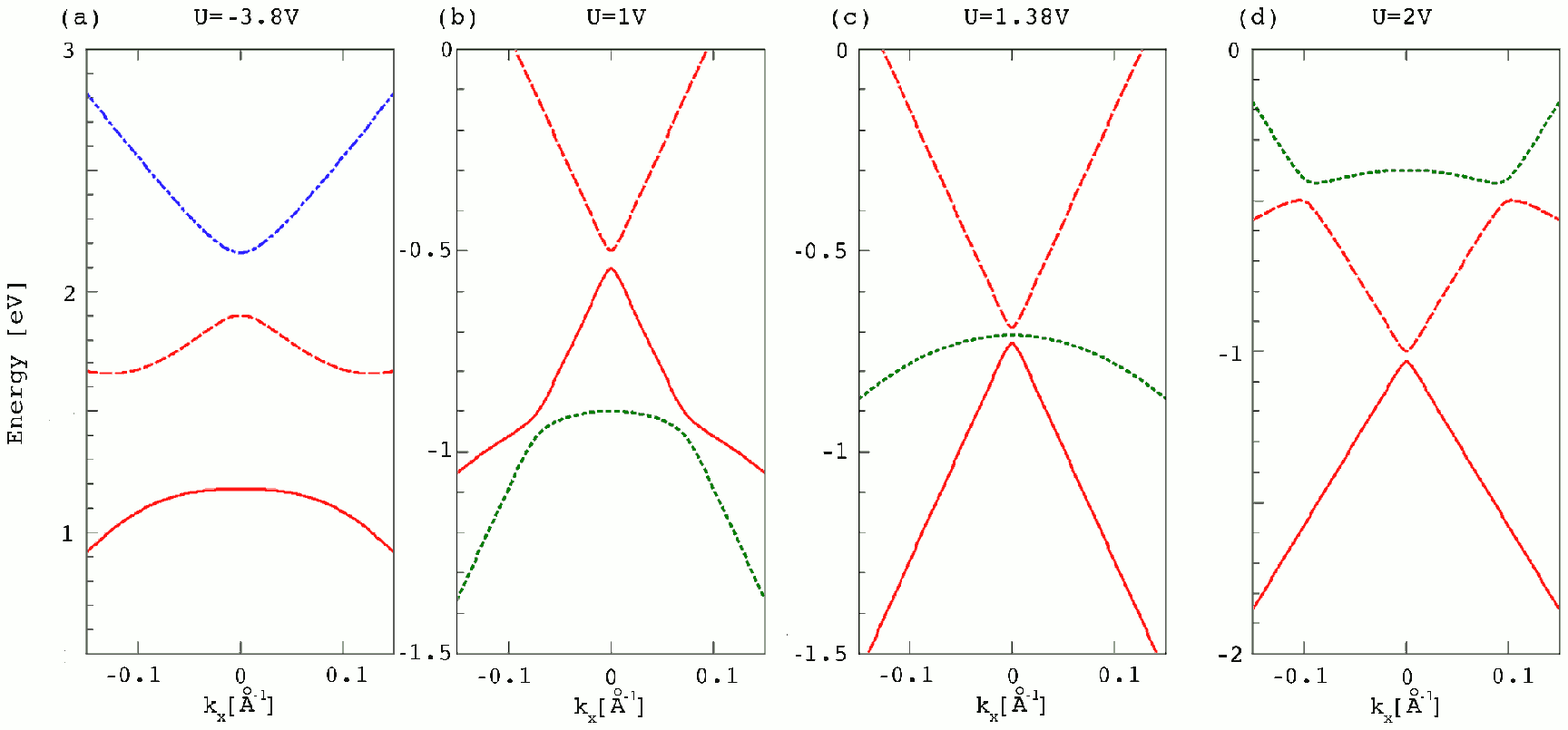}
\caption{\label{napiecia}(Color online) Band structures of graphene on h-BN bilayer for different values of $U$. Blue dotted-dashed line denotes a band with the value $E_{1}$ (eq.\ref{rozw1}) at K point; red dashed line is assigned to the value $E_{2}$ (eq. \ref{rozw2}); olive dotted to the value $E_{4}$ (eq. \ref{rozw4}) and red solid to the value $E_{3}$ (eq.\ref{rozw3}). The TB parameters are taken from Table \ref{parametry}.}
\end{figure*}

The electronic properties of a graphene bilayer in the presence of an external electric field applied perpendicularly to the system have been widely studied.\cite{mccann,biased} We are interested here in the properties of bilayer system consisting of graphene and h-BN monolayers. The external perpendicular field gives rise to an electrostatic energy difference between these two layers which is parametrized by $U$. The interlayer asymmetry induced by an external field in the bilayer system shown in Fig. \ref{sieci}\,(c) can be described by the hamiltonian\cite{biased} in the tight binding approximation:
\begin{equation}H=\left( \begin{array}{cccc}\vspace{0.1in}
\epsilon^{\alpha_{1}}_{2p_{z}}+\frac{U}{2}& \gamma_{0}'f(\vec{k})&0&\gamma_{1}\\\vspace{0.1in}
\gamma_{0}'f(\vec{k})^{*} & \epsilon^{\beta_{1}}_{2p_{z}}+\frac{U}{2}&0&0\\\vspace{0.1in}
0& 0&\epsilon^{\alpha_{2}}_{2p_{z}}-\frac{U}{2}&\gamma_{0}f(\vec{k})\\
\gamma_{1}& 0&\gamma_{0}f(\vec{k})^{*}&\epsilon^{\beta_{2}}_{2p_{z}}-\frac{U}{2}\end{array} \right)\end{equation}

When the condition (\ref{warunek}) holds, one can receive the eigenvalues for arbitrary parameters:
\begin{equation}E_{1}=\frac{1}{2}\epsilon^{\alpha_{1}}_{2p_{z}}+\sqrt{\biggl(\frac{\epsilon^{\alpha_{1}}_{2p_{z}}}{2}\biggr)^{2}+\gamma_{1}^{2}+\frac{U}{2}(\epsilon^{\alpha_{1}}_{2p_{z}}+\frac{U}{2})}\label{rozw1}\end{equation}\begin{equation}E_{2}=-\frac{U}{2}\label{rozw2}\end{equation}\begin{equation}E_{3}=\frac{1}{2}\epsilon^{\alpha_{1}}_{2p_{z}}-\sqrt{\biggl(\frac{\epsilon^{\alpha_{1}}_{2p_{z}}}{2}\biggr)^{2}+\gamma_{1}^{2}+\frac{U}{2}(\epsilon^{\alpha_{1}}_{2p_{z}}+\frac{U}{2})}\label{rozw3}\end{equation}\begin{equation}E_{4}=\epsilon^{\beta_{1}}_{2p_{z}}+\frac{U}{2}\label{rozw4}\end{equation}
where we have assumed, for simplicity, $\epsilon^{\alpha_{2}}_{2p_{z}}=\epsilon^{\beta_{2}}_{2p_{z}}=0$.

As a starting point to study the influence of the electric field on the band structure we find the values of parameters $U$ for which two bands touches in the $K$ ($K'$) point. A simple analysis shows that it occurs in three cases. First one gives the value of $U$ when the bands corresponding to $E_{1}$ and $E_{3}$ are joined:
\begin{equation}U_{A}=-\epsilon^{\alpha_{1}}\pm 2 i \gamma_{1}\label{EA}\end{equation}
second one when the bands corresponding to $E_{3}$ and $E_{4}$ meet in K point:
\begin{equation}U_{B}=\frac{(\epsilon^{\beta_{1}}_{2p_{z}})^{2}-\epsilon^{\alpha_{1}}_{2p_{z}}\epsilon^{\beta_{1}}_{2p_{z}}-\gamma_{1}^{2}}{\epsilon^{\alpha_{1}}_{2p_{z}}-\epsilon^{\beta_{1}}_{2p_{z}}}\label{EB}\end{equation}
and the third one comes from the condition $E_{2}$=$E_{4}$:
\begin{equation}U_{C}=-\epsilon^{\beta_{1}}\label{EC}\end{equation}

Figure \ref{charakterystyka} shows the changes of a band gap as a function of $U$ while the insets present the band structure around the $K$ point in three cases described by conditions (\ref{EA}), (\ref{EB}), (\ref{EC}) and indicated by letters A, B and C on the plot. Since the condition (\ref{EA}) gives an imaginary value of $U_{A}$, the band structure in point A on the curve can be considered only by taking its real part. Inset A in Fig.~\ref{charakterystyka} depicts the band structure determined for $U_{A}=-\epsilon^{\alpha_{1}}$ and it follows that it is the lowest value of the voltage, for which a direct band gap is observed. It achieves a value of $\gamma_{1}$. The second condition (\ref{EB}) predicts that two valence bands touches, which is shown in inset B of Fig.~\ref{charakterystyka}. Finally, condition (\ref{EC}) leads to the situation when the valence and conduction bands meet in the corner of the Brillouin zone (inset C in Fig. \ref{charakterystyka}). Moreover, in the last case we observe a large slope of conduction band in the vicinity of $K$ point which indicates that nearly conical touching can be recognized ($m^{*}=3.4\cdot 10^{-3}\,m_{\textrm{e}}$).

In Fig. \ref{napiecia} we present the band structures for the values of voltages chosen in the way to show what happen in the regions outside the boundaries determined by $-\epsilon^{\alpha_{1}}$ and $-\epsilon^{\beta_{1}}$ and in the regions between points A - B and B - C (see Fig. \ref{charakterystyka}). For the values of voltages lower than $-\epsilon^{\alpha_{1}}$ (Fig. \ref{napiecia}\,(a)) and higher than $-\epsilon^{\beta_{1}}$ (Fig. \ref{napiecia}\,(d)) the band gap is indirect which is not observed in case of the bilayer graphene when valence and conduction bands have the same Mexican hat shape. Moreover, for higher positive voltage (Fig. \ref{napiecia}\,(d)) we observe the restructuring of conduction band which seems to be a result of an anticrossing interaction between highly localized states of electronegative nitrogen atoms with the extended conduction band states of graphene.\cite{walukiewiczunp} The electronic states loose their localized character and form band through an interaction with extended states. 
For the voltages from A - B range of values (Fig. \ref{napiecia}\,(b)), we observe an upward movement of the valence bands edges. It can be interpreted as induced by hybridization of the carbon $p_{z}$ states comprising the higher valence band with the close lying localized $p_{z}$ states of N. This bands anticrossing phenomenon leads to the shift of the valence bands edges and to the reduction of the band gap.\cite{walukiewiczprb}

It is worthwhile to notice that the band structure modifications of the bilayer graphene/h-BN system induced by the external electric field is similar to the behaviour of the band structure in mismatched III-N-V and III-V semiconductor alloys.\cite{walukiewiczprb, walukiewiczprl, walukiewiczsol}

\section{Density functional calculations}
We have performed density functional calculations for bilayer graphene/h-BN system using PWscf package\cite{pwscf} of the Quantum Espresso distribution\cite{qe}. A plane wave expansion with an energy cutoff at 400 Ry and the Perdew-Zunger LDA exchange~-~correlation functional\cite{perdewzunger} were used. Core electrons were described by norm conserving pseudopotentials. The Brillouin zone was sampled on a uniform 36 $\times$ 36 $\times$ 1 grid with the tetrahedron method\cite{tetrahedron}. The interactions with spurious replicas along the z direction were avoided by setting a large size of a supercell $z\sim16$\,\AA. We have taken a value of lattice constant a=2.456\,\AA. Total energies were convergent to $10^{-4}$~Ry. 

\begin{figure}
\includegraphics{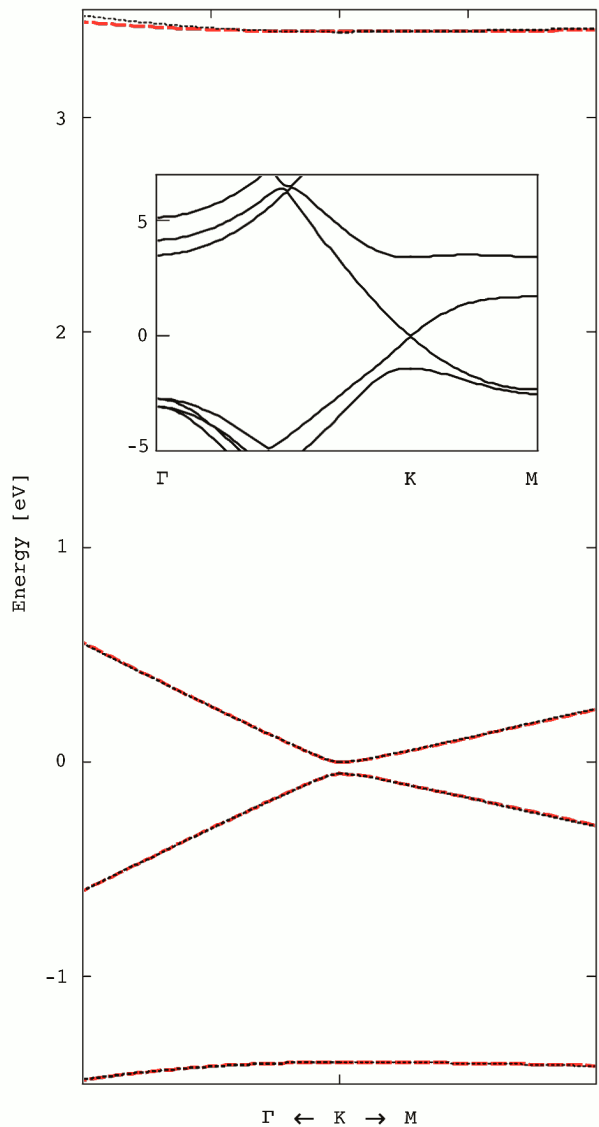}
\caption{\label{dopasowanie}(Color online) The TB fit along $\Gamma KM$ line in the vicinity of K point for LDA calculations. Red dashed lines denote DFT bands while black dotted lines denote TB fit. The inset show DFT band structure in $\Gamma K$ and KM directions. The energies are measured from the zero level fixed by the condition $\epsilon^{\alpha_{2}}_{2p_{z}}\,=\,0$.}
\end{figure}

\begin{table}[b]
\caption{\label{parametry}Tight binding parameters calculated from best fit to DFT data. All values are given in electronvolts. The energies $\epsilon^{\alpha_{1}}_{2p_{z}}$ and $\epsilon^{\beta_{1}}_{2p_{z}}$ are measured from the zero level fixed by the condition $\epsilon^{\alpha_{2}}_{2p_{z}}\,=\,0$.}
\begin{ruledtabular}
\begin{tabular}{ccccc}
$\gamma_{0}$&$\gamma_{0}'$&
$\gamma_{1}$&$\epsilon^{\alpha_{1}}_{2p_{z}}$&$\epsilon^{\beta_{1}}_{2p_{z}}$\\
\hline
2.64&2.79&0.43&3.34&-1.40\\
\end{tabular}
\end{ruledtabular}
\end{table} 

\begin{figure}
\includegraphics{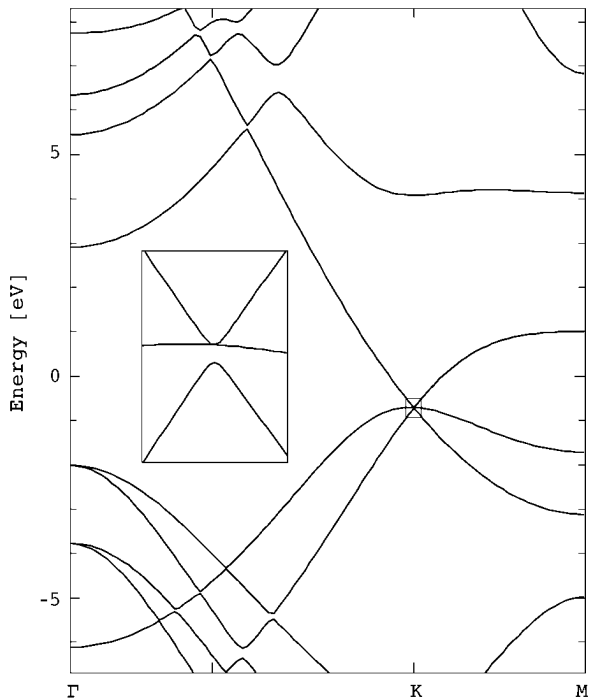}
\caption{\label{dft1}DFT band structure along $\Gamma KM$ line in momentum space of bilayer graphene/h-BN in the presence of external electric field with the effective value of U = 1.40 V applied in the direction perpendicular to the layers. The inset presents zoom of the band structure in the vicinity of K point, where the band gap is closed.}
\end{figure}

We fit the TB band structure with DFT data to test the model analyzed in Sec.~II\,B. The fitting procedure has been applied to the points taken from the straight vicinity of K point along $\Gamma K$ line using the least squares method. We present the calculated DFT and TB band structures in Fig. \ref{dopasowanie}. It is easy to see, that both electronic structures are well fitted. The small discrepancies start to appear only at the higher conduction band. The obtained values of TB parameters are listed in Table \ref{parametry}.

As it is expected, in local density approximation (LDA), the difference $|E_{1}-E_{4}|$ between boron nitride bands is underestimated. Its value should be higher than 5.9 eV\cite{gwhbn}, while the currently determined value is in the order of 4.7 eV which is in a good agreement with LDA results for pure boron nitride honeycomb\cite{turcy}. 

In the second stage, we have performed the \textit{ab initio} calculations for graphene/h-BN system in the presence of a constant electric field simulated by a sawlike potential. The results confirm TB predictions related to the band structure modifications described in Sec.~II$\,$c. It should be noticed that similarly to the case of bilayer graphene\cite{hongkimin,gava} we observe screening effects. It means that in order to obtain expected shift and deformation of bands we have to apply an electric field higher than estimated in TB model. Figure \ref{dft1} presents DFT band structure for the case given in inset C of Fig. \ref{charakterystyka}. The TB predicted value of $U=-\epsilon^{\beta_{1}}$ is too low to reproduce the gap closing, in fact we have to apply $U=4.15$~V.

\begin{figure}
\includegraphics{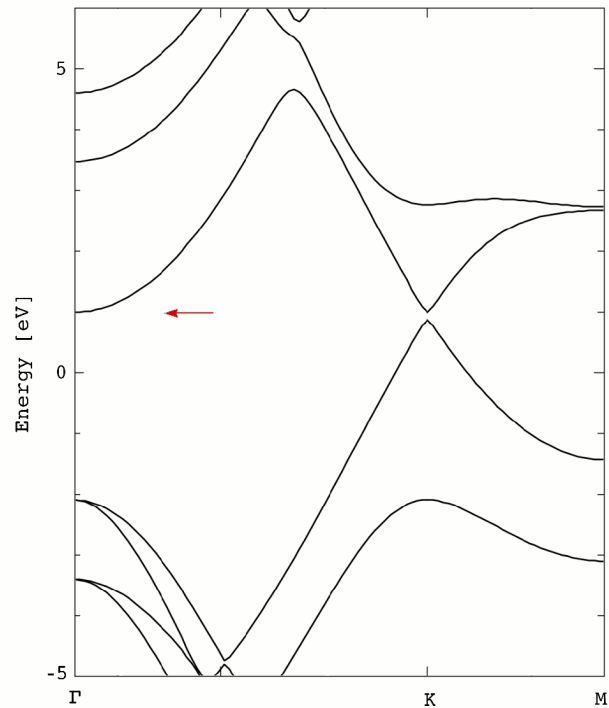}
\caption{\label{dft2}(Color online) DFT band structure along $\Gamma KM$ line in momentum space of bilayer graphene/h-BN  in the presence of external electric field with the effective value of U = -2.0 V applied in the direction perpendicular to the layers. Red arrow points out the edge of conduction band originating from hybridized states comprising $\pi$ carbon band and $\sigma$ boron nitride band.}
\end{figure}

In Fig. \ref{dft2} we show the band structure of the graphene on h-BN in the presence of an electric field parametrized by negative U of the effective value equal to -2 V. We can observe a direct gap of over 127 meV between valence and conduction band. One can notice, however, that at the $\Gamma$ point one of the boron nitride originating bands starts to coincide with the energy level of the bottom edge of carbon conduction band (at the K point), which leads to an indirect band gap of the system. For the higher absolute values of potential difference, a band gap decreases due to the presence of unwanted edge and achieves negative values typical for semimetals. When a system gains semimetallic character, very strong screening effects prevent further increase of the effective voltage. 

\section{Final remarks}
We have studied the electronic properties of a graphene/h-BN bilayer system using the minimal tight binding model. Particular attention has been given to the presence of external electric field perpendicular to the bilayer system which gives rise to a finite gap in the spectrum, whose opening and width are controlled by the applied voltage. The use of TB approach allows to show that graphene/h-BN system and bilayer graphene are very closely related. 

Moreover, we propose a very simple model as a starting point for similar considerations of more sophisticated substrates and interactions with metallic contacts.\cite{holendrzy, klusek} Further generalization on a larger number of AB stacked layers of graphene and AA stacked h-BN sheets can also be made.

We have shown that the band structure may be modified and tuned to meet specific requirements for potential logic devices. We report the energy gap tuning in the range starting from 0 to about 130 meV and effective masses in conduction band varying from $3.4\cdot 10^{-3}\,m_{\textrm{e}}$ to $1.1\cdot 10^{-2}\,m_{\textrm{e}}$, which are smaller than Dirac fermions' masses in bilayer graphene. One can easily predict that additional quasi-particle interactions included in calculations, for instance within a GW scheme, will indicate an increased value of the gap.\cite{hbn} Moreover, DFT as a ground-state theory does not give a proper description of excited states. Therefore, it is possible that an energy of bottom edge of the conduction band (see Fig.~\ref{dft2}) is not correctly determined. On the other hand, if one could engineer the band structure to shift or deform this conduction band edge then the voltage induced band gap could be as large as predicted by TB calculations (over 430 meV see Fig.~\ref{napiecia}\,(a)).

Since a significant step towards the bilayer graphene FET production has been recently taken,\cite{nanoletters} it has to be stressed that various modified bilayer systems could play a crucial role in the future nanoelectronics. Recent studies on transferring exfoliated graphene onto arbitrary substrates\cite{nakladanie} suggest that the fabrication of graphene on~h-BN could be now realized.

\begin{acknowledgments}
We are grateful to Piotr Kosi\'{n}ski for helpful discussions and to Igor W\l asny for careful reading.

This work is financially supported by Polish Ministry of Science and Higher Education in the frame of grant "Investigations of electronic structure of graphene deposited on conductive and nonconductive surfaces by STM/STS/CITS/AFM and quantum electrodynamics N~N202~204737". One of us (J.S.) acknowledges support of the European Social Fund and Budget of State implemented under the Integrated Regional Operational Program, Project: Scholarship supporting postgraduate students' innovative scientific research.

Figure 1 was prepared using the XCrysDen program.\cite{kokalj}
\end{acknowledgments}
\appendix*
\providecommand{\noopsort}[1]{}\providecommand{\singleletter}[1]{#1}%

\end{document}